\documentclass[aps, prx, 10pt, amssymb,amsmath,superscriptaddress,tightenlines,twocolumn,notitlepage,longbibliography] {revtex4-2}
\usepackage{graphicx}
\usepackage{amsmath}
\usepackage{amssymb}
\usepackage{bm}
\usepackage{color}
\usepackage{hyperref}
\usepackage{tabularx}
\usepackage{multirow}
\usepackage{booktabs}
\usepackage{braket}
\usepackage{epstopdf}
\usepackage{pdfpages}
\usepackage{pgffor}

\makeatletter
\AtBeginDocument{\let\LS@rot\@undefined}
\makeatother

\def\supplementfilename{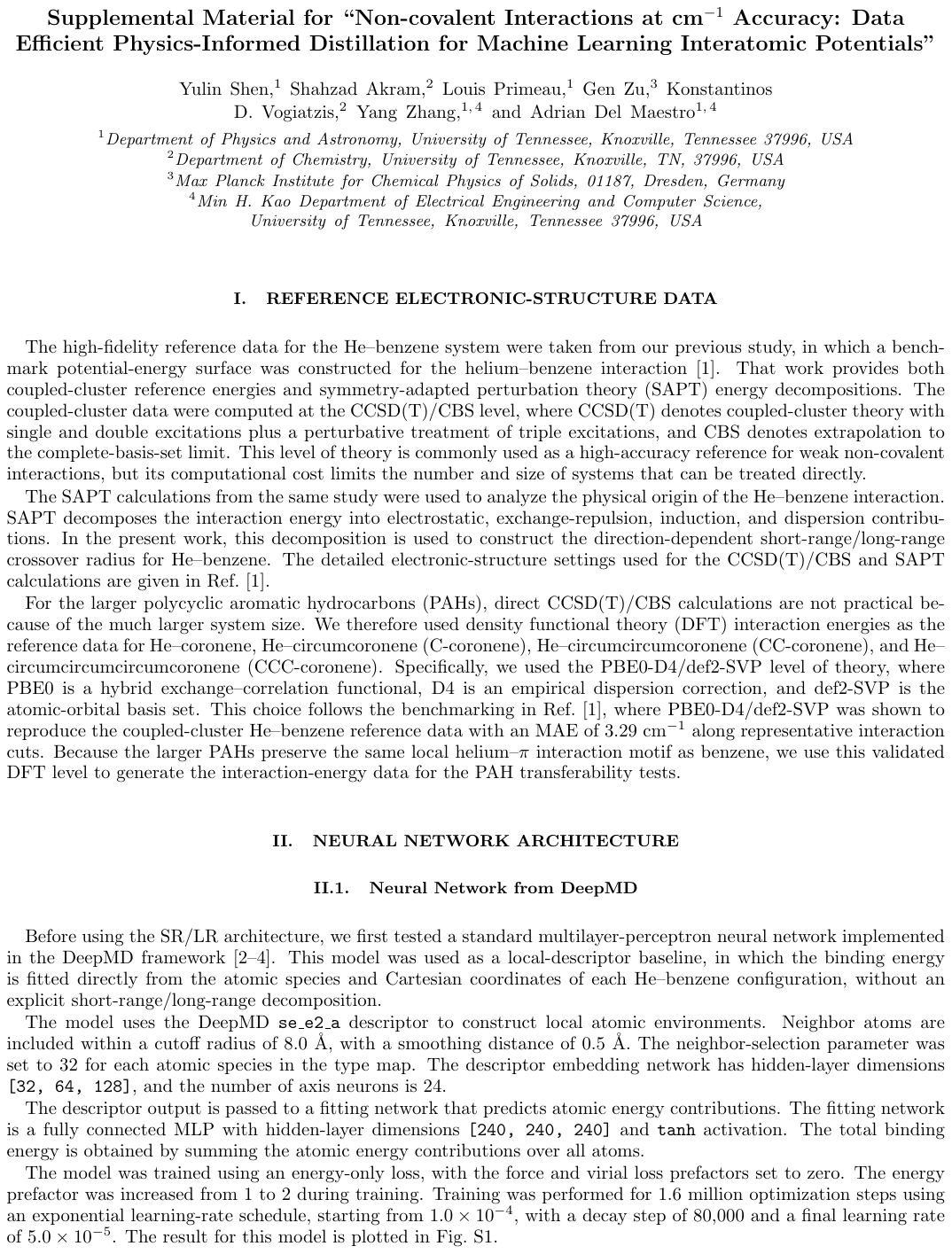}

\pdfximage{\supplementfilename}
\def\numbersupplementpages{\the\pdflastximagepages}

\newif\ifarXiv
\arXivtrue

\usepackage[color=orange!60,textsize=scriptsize]{todonotes}
\usepackage{siunitx}
\DeclareSIUnit\angstrom{\text{\AA}}
\makeatletter
\DeclareRobustCommand{\r}{\ifmmode\expandafter\mathring\else\expandafter\@ring\fi}
\makeatother

\begin{document}

\title{Non-covalent Interactions at cm$^{-1}$ Accuracy: Data Efficient Physics-Informed Distillation for Machine Learning Interatomic Potentials}

\author{Yulin Shen}
\affiliation{Department of Physics and Astronomy, University of Tennessee, Knoxville, Tennessee 37996, USA}

\author{Shahzad Akram}
\affiliation{Department of Chemistry, University of Tennessee, Knoxville, Tennessee 37996, USA}

\author{Louis Primeau}
\affiliation{Department of Physics and Astronomy, University of Tennessee, Knoxville, Tennessee 37996, USA}

\author{Gen Zu}
\affiliation{Max Planck Institute for Chemical Physics of Solids, 01187, Dresden, Germany}

\author{Konstantinos D. Vogiatzis}
\affiliation{Department of Chemistry, University of Tennessee, Knoxville, Tennessee 37996, USA}

\author{Yang Zhang}
\affiliation{Department of Physics and Astronomy, University of Tennessee, Knoxville, Tennessee 37996, USA}
\affiliation{Min H. Kao Department of Electrical Engineering and Computer Science, University of Tennessee, Knoxville, Tennessee 37996, USA}

\author{Adrian Del Maestro}
\affiliation{Department of Physics and Astronomy, University of Tennessee, Knoxville, Tennessee 37996, USA}
\affiliation{Min H. Kao Department of Electrical Engineering and Computer Science, University of Tennessee, Knoxville, Tennessee 37996, USA}

\begin{abstract}
Foundation models in atomistic machine learning encode interaction physics across diverse atomic environments, but whether that structure can be transferred when building specialist potentials at quantum-chemical accuracy remains open. Here we show that knowledge distillation from a pretrained universal machine-learning interatomic potential (MLIP), followed by coupled-cluster fine-tuning with single and double excitations and perturbative triples [CCSD(T)], transfers not only low-cost labels but a physically meaningful prior on interaction length scales, anisotropy, and the repulsive--dispersive balance, which CCSD(T) data then sharpens to quantum-chemical accuracy. For He--benzene, fine-tuning with $30\%$ of the CCSD(T) data outperforms direct training using the full $80\%$---a $\sim 63\%$ reduction in the high-fidelity compute budget. A symmetry-adapted perturbation theory (SAPT)-informed adaptive short-range/long-range architecture further lowers the validation MAE from \SI{0.75}{\centi\meter^{-1}} to \SI{0.49}{\centi\meter^{-1}}. Across a circumarene series of polycyclic aromatic hydrocarbons (PAHs), swapping the MLIP teacher under an otherwise identical pipeline changes the coronene error by an order of magnitude while leaving the larger PAHs stable---direct evidence that distillation transfers physical structure, not labels alone. Together, these results identify the choice of pretrained teacher as a primary design axis for data-efficient quantum-chemical-accuracy potentials, alongside architecture and training protocol.
\end{abstract}

\maketitle
\section{Introduction}
Describing non-covalent intermolecular interactions at quantum accuracy is a central challenge of atomistic modeling, because subtle energy differences on the order of cm$^{-1}$ can govern adsorption geometries, spectroscopic signatures, and molecular recognition~\cite{hobza2016introduction,muller2000noncovalent,sherrill2013energy}. The coupled-cluster method with single and double excitations and perturbative triples [CCSD(T)]~\cite{purvis1982full, raghavachari1989fifth}, extrapolated to the complete-basis-set limit (CBS), provides a standard high-accuracy reference for weak non-covalent interactions. Although CCSD(T)/CBS is widely regarded as a gold-standard description of such interactions, its cost is prohibitive for large training sets. Canonical CCSD scales as $\mathcal{O}(N^6)$ (where $N$ is the number of basis functions) and the perturbative triples correction scales as $\mathcal{O}(N^7)$ with system size, while CBS extrapolation requires calculations in multiple basis sets. As a result, practical high-fidelity reference datasets are typically limited to at most thousands of configurations, far below what is needed to train an accurate neural-network interatomic potential without a pretrained starting point. Machine learning force fields offer a route around this bottleneck by delivering near-\textit{ab initio} accuracy at a fraction of the cost of electronic structure methods~\cite{behler2007generalized, rupp2012fast,schutt2017schnet,batzner20223,unke2021machine}, and general-purpose machine-learning interatomic potentials (MLIPs) in particular combine broad chemical coverage, transferable representations, and near-\textit{ab initio} efficiency within a single pretrained model family. Representative examples include MACE~\cite{batatia2022mace,batatia2025foundation}, MatGL/M3GNet~\cite{chen2022universal,deng2023chgnet}, MatterSim~\cite{yang2024mattersim}, and the Orb family~\cite{neumann2024orb}. Together, these models show that pretrained atomistic models can serve as useful starting points for building potentials tailored to specific systems and levels of theory.

Broad coverage, however, does not by itself guarantee the accuracy needed for a particular physical problem. This limitation becomes especially severe for weakly bound noncovalent systems, where shallow and strongly anisotropic interaction landscapes demand precision well beyond what general-purpose MLIPs achieve in their pretrained form. Adapting such models to a specialized target through frozen transfer, fine-tuning, or related approaches has accordingly attracted growing attention~\cite{chen2022machine, smith2019approaching, zhang2026constructing, wang2025pre,zaverkin2023transfer}. Yet such adaptation is usually treated as an optimization strategy, including warm-start initialization, parameter-efficient transfer~\cite{hu2022lora}, or data-efficient fine-tuning, rather than as a mechanism for refining interaction patterns inherited from a pretrained representation. A natural question then arises: can the physical prior carried by an MLIP be used to reach quantum-chemical accuracy on a specific target interaction with only a minimal amount of expensive high-fidelity data?

In this work, we develop a hybrid distillation and fine-tuning framework that reaches quantum-chemical accuracy on weak intermolecular interactions with minimal high-fidelity training data. We first use a pretrained MLIP to label many target-relevant configurations, so that the specialized model learns the coarse structure of the interaction surface, including its length scale, anisotropy, and attractive--repulsive balance. CCSD(T) fine-tuning then corrects this representation toward the target level of theory. We further introduce an adaptive short-range (SR)/long-range (LR) architecture informed by symmetry-adapted perturbation theory (SAPT)~\cite{jeziorski1994perturbation,hohenstein2012wavefunction,parrish2017psi4}, which uses the geometry-dependent crossover between short-range repulsion and long-range dispersion to define an adaptive short-range cutoff. We demonstrate this strategy on a weakly bound benchmark He--benzene system, which exhibits a strongly anisotropic, geometry-dependent energy landscape, and on a related series of polycyclic adsorbates with increasing numbers of carbon rings. Together, these molecular systems provide a demanding test of both low-data efficiency and transferability. Overall, these results indicate that hybrid MLIP--CCSD(T) adaptation, combined with a physically informed SR/LR architecture, can reach sub-cm$^{-1}$ accuracy for He--benzene and may provide a data-efficient route for constructing potentials for weak intermolecular interactions.

\section{Teacher-Guided Distillation and High-Fidelity Adaptation}

\begin{figure}[h]
    \centering
    \includegraphics[width=1\linewidth]{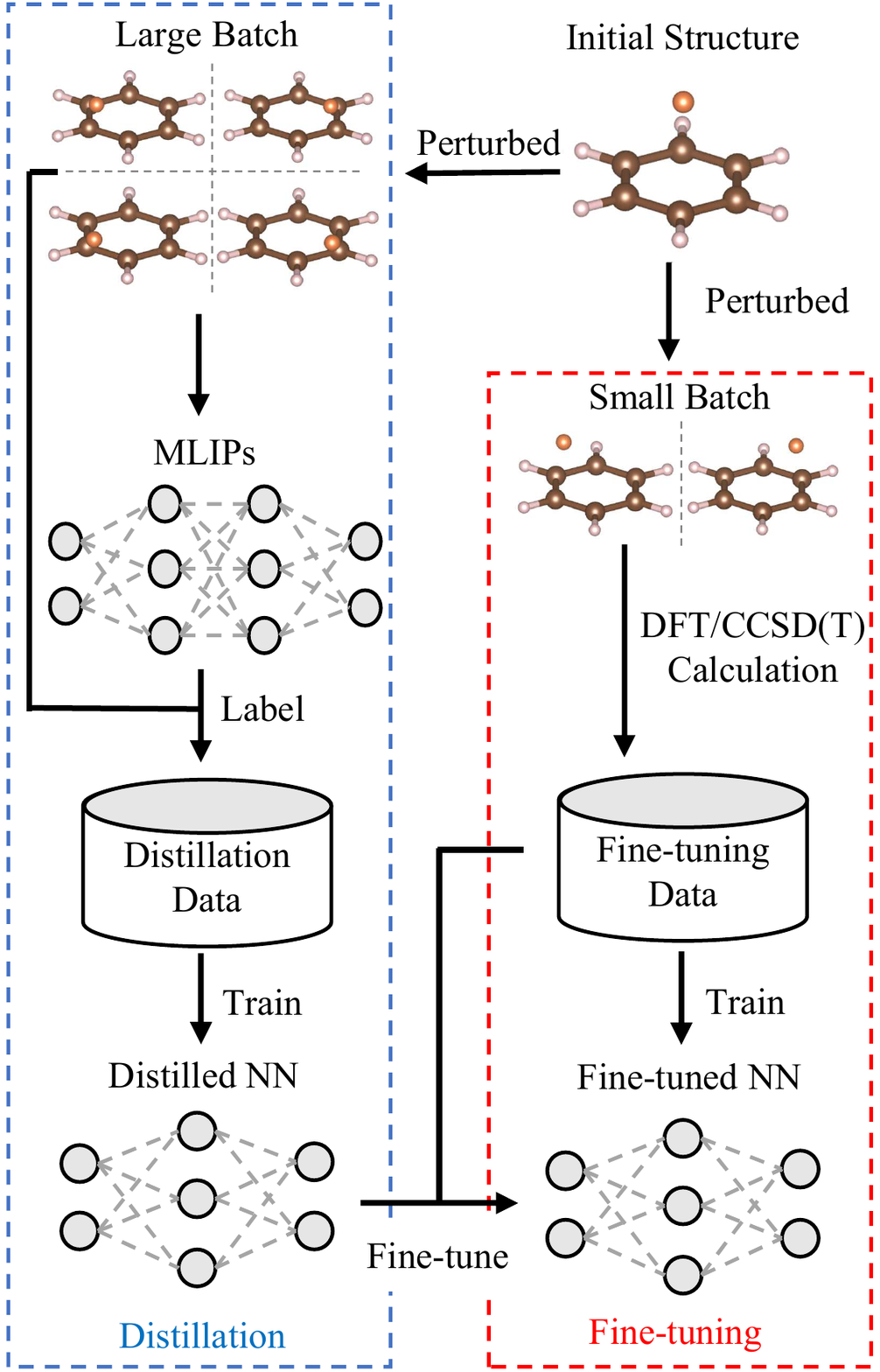}
    \caption{{Teacher-guided universal-to-specialized adaptation framework.} Starting from an initial He--benzene structure, two sets of perturbed configurations are generated. A large batch is labeled by a pretrained general-purpose machine-learning interatomic potential (MLIP) and used to train a distilled student neural network (blue box left). A separate smaller batch is evaluated with high-fidelity DFT or CCSD(T) calculations and used to fine-tune the distilled student to the target level of theory (red box right). The fine-tuned model is initialized from the distilled student rather than from a randomly initialized network.}
    \label{workflow}
\end{figure}

We consider distillation and fine-tuning as a teacher-guided route for universal-to-specialized adaptation in atomistic learning, as illustrated in Fig.~\ref{workflow}. 
This workflow is not specific to He--benzene, but provides a general strategy for adapting pretrained MLIPs to a target interaction class whenever target-relevant configurations can be sampled and a smaller set of high-fidelity reference data is available. In this setting, a pretrained MLIP labels a large set of target-relevant configurations at low computational cost. Although these labels are only energies, they define an approximate interaction surface learned from the pretrained model, including its length scale, anisotropy, and attractive--repulsive balance. A lightweight student model is first trained on this teacher-labeled surface~\cite{hinton2015distilling,bucilua2006model,gou2021knowledge}, and is then fine-tuned with a smaller set of high-fidelity reference data to correct the surface to the target level of theory. Unlike simple warm-start initialization or parameter-efficient transfer strategies~\cite{pan2009survey,hu2022lora,chen2022machine,zaverkin2023transfer}, the present approach first trains the student on a large set of teacher-labeled, target-relevant configurations before high-fidelity fine-tuning, in the spirit of teacher--student distillation~\cite{hinton2015distilling}. The teacher-labeled configurations therefore provide an approximate interaction surface that is subsequently refined with DFT or CCSD(T) reference data. Because this starting point depends on the teacher, different MLIPs do not give equivalent downstream performance; the outcome can depend on the match between the teacher, the student architecture, the amount of fine-tuning data, and the target interaction class~\cite{vandermause2020fly,schran2020committee,gardner2024synthetic}. We therefore use He--benzene as a stringent weak-interaction benchmark to compare three teacher-labeling routes before the high-fidelity fine-tuning stage, as summarized in Table~\ref{table1}.  Tests on additional models are included in the Supplemental Material~\cite{supplemental}. 
\begin{table}[h!tbp]
\centering
\caption{{Validation errors (in \si{\centi\meter^{-1}}) of distilled He--benzene student models before CCSD(T) fine-tuning.} In order to select a teacher model, each student is trained using labels generated by a different teacher-labeling route: direct DFT, \texttt{MatterSim}, or \texttt{Orb}. The MAEs are evaluated against the CCSD(T) reference energies, so the numbers measure the distilled student models rather than the bare teacher predictions. The \texttt{Orb} teacher yields the lowest error and is our focus in subsequent experiments.}
\label{table1}
\begin{tabular}{c c c c}
\toprule
 & DFT & \texttt{MatterSim} & \texttt{Orb} \\
\midrule
MAE  & 96.61 & 316.61 & 50.55 \\
RMSE  & 226.96 & 367.81 & 70.16 \\
MAX error  & 2108.97 & 993.84 & 584.83 \\
\bottomrule
\end{tabular}
\end{table}

\section{Physically Informed Short-Range/Long-Range Model}

To demonstrate our teacher-guided adaptation framework, we apply it to He--benzene, a well-studied prototype for helium adsorption on graphitic surfaces and related polycyclic aromatic hydrocarbons (PAHs) that is simultaneously accessible to CCSD(T)/CBS reference calculations~\cite{akram2026,shirkov2024ab,cappelletti2002molecular,brupbacher1994intermolecular,schiller2021adsorption}. Its interaction landscape is weak and strongly anisotropic, with a geometry-dependent crossover between short-range exchange repulsion and long-range dispersion-dominated attraction. Accurately modeling this crossover requires an adaptive SR/LR architecture with a SAPT-informed adaptive cutoff network for the anisotropic SR boundary; in what follows, we compare such an architecture with its fixed-cutoff counterpart. The specialized model needs to describe how the interaction is divided between the SR and LR regimes, and we start by decomposing the total interaction energy as the sum of short-range and long-range contributions~\cite{tang1984improved,jeziorski1994perturbation,hohenstein2012wavefunction,parrish2017psi4},
\begin{equation}
E_{\mathrm{tot}}(\mathbf{R}) = E_{\mathrm{SR}}(\mathbf{R}) + E_{\mathrm{LR}}(\mathbf{R}),
\label{eq:Etot}
\end{equation}
where $\mathbf{R}$ denotes the atomic coordinates. For He--benzene, the main difficulty is that the SR/LR boundary is not isotropic; it depends on the He approach direction relative to the benzene molecular frame (motivating the adaptive cutoff introduced below).

We first construct a fixed-cutoff SR/LR model as a direct implementation of Eq.~\eqref{eq:Etot} with a single global SR cutoff. The input is constructed from the set of relative vectors $\{\mathbf{r}_{\mathrm{He},i}\}$ between the He atom and each atom (C, H) in the benzene molecule, with distances expanded in compact-support radial basis functions. The SR branch, defined here as the network component used to model $E_{\mathrm{SR}}$, maps these descriptors to atom-wise energy contributions through a multilayer perceptron (MLP) and sums them to obtain $E_{\mathrm{SR}}$. The LR branch is the corresponding network component used to model $E_{\mathrm{LR}}$: it uses a separate descriptor with a fixed cutoff, and a latent MLP generates coefficients that are passed to a SOGNet-based~\cite{ji2025machine} module to evaluate $E_{\mathrm{LR}}$. This fixed-cutoff architecture provides the simplest SR/LR decomposition, but it cannot follow the direction-dependent SR/LR crossover near the He--benzene binding well.

\begin{figure}[h]
    \centering
    \includegraphics[width=0.99\linewidth]{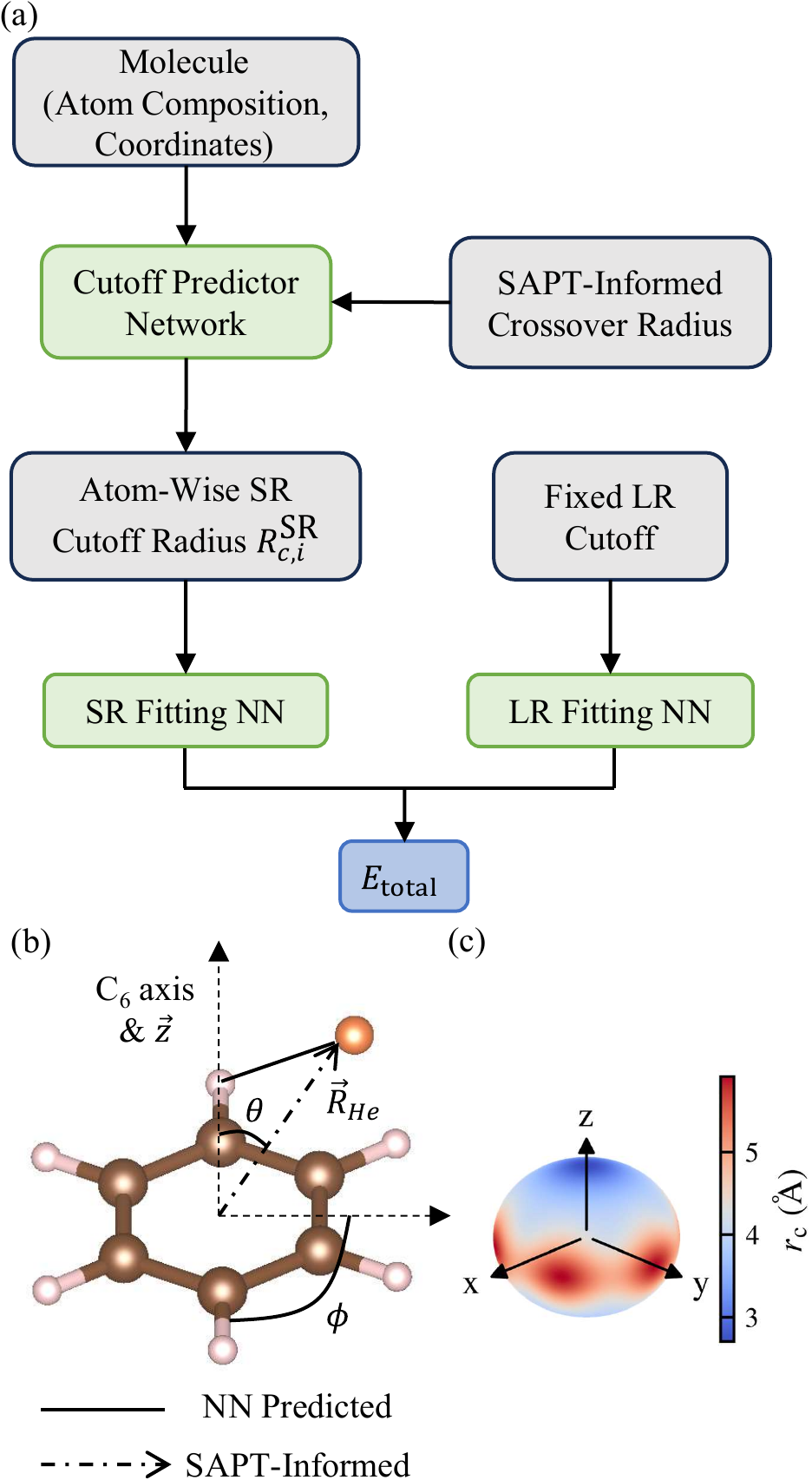}
    \caption{{SAPT-informed architecture and cutoff construction.}
    (a) Adaptive SR/LR model for He--benzene, with a SAPT-informed SR cutoff and fixed LR cutoff. 
    (b) The center-based SAPT crossover radius $R_c^{\mathrm{SAPT}}(\Omega)$ is defined by the He direction relative to the benzene center and mapped to atom-wise SR cutoffs $R_{c,i}^{\mathrm{SR}}$.
    (c) The fitted $R_c^{\mathrm{SAPT}}(\Omega)$ varies from approximately $2.7$ to $5.8~\mathrm{\AA}$, motivating a geometry-adaptive SR cutoff.}
    \label{architecture}
\end{figure}

To address this limitation, we introduce a SAPT-informed adaptive SR/LR architecture, shown schematically in Fig.~\ref{architecture}.

Here, SAPT provides a physically motivated definition of the SR cutoff, rather than merely an additional fitting target or input feature. For each He position, SAPT provides a direction-dependent crossover radius $R_c^{\mathrm{SAPT}}(\Omega)$ referenced to the benzene molecular center, defined here as the geometric center of the six carbon atoms. Here, $\Omega=(\theta,\phi)$ denotes the direction of the vector from the benzene molecular center to the He atom in the benzene-fixed coordinate frame. The polar angle $\theta$ is measured from the molecular $C_6$ axis normal to the benzene plane, and the azimuthal angle $\phi$ is measured in the benzene plane from the fixed $x$ axis as shown in Fig.~\ref{architecture}(b). As shown in Fig.~\ref{architecture}(c), the fitted SAPT cutoff varies strongly with angle, from roughly $2.7$ to $5.8~\mathrm{\AA}$, so a single fixed SR cutoff would either overextend the SR descriptor in some directions or truncate it too early in others. As illustrated in Fig.~\ref{architecture}(b), this SAPT quantity is defined relative to the molecular center and therefore does not directly provide the pair-specific cutoff needed for the SR descriptor. We therefore map it to atom-wise SR threshold through a cutoff predictor network,
\begin{equation}
R_{c,i}^{\mathrm{SR}} = g_{\boldsymbol{\eta}}\!\left(Z_i, r_{\mathrm{He},i}, \hat{\mathbf{r}}_{\mathrm{He},i}, R_c^{\mathrm{SAPT}}(\Omega)\right).
\end{equation}
Here $g_{\boldsymbol{\eta}}$ is a trainable cutoff-prediction network with parameters $\boldsymbol{\eta}$. For atom $i$, $Z_i$ denotes the atomic species, $r_{\mathrm{He},i}=|\mathbf{r}_{\mathrm{He},i}|$ is the He--atom distance, and $\hat{\mathbf{r}}_{\mathrm{He},i}$ is the corresponding unit vector. The output $R_{c,i}^{\mathrm{SR}}$ is a pair-specific cutoff for the SR descriptor associated with the He--atom-$i$ pair.

These cutoffs set the compact support of the SR descriptor separately for each He--benzene atom pair, rather than applying a single global SR cutoff to all pairs. During training, we include two regularization terms: one penalizes the variance of $R_{c,i}^{\mathrm{SR}}$ among symmetry-equivalent C or H atoms, and the other keeps the C- and H-averaged predicted cutoffs close to the center-based SAPT value $R_c^{\mathrm{SAPT}}(\Omega)$. Details of the SAPT cutoff construction, spherical-harmonic fitting, and regularization terms are given in the Supplemental Material~\cite{supplemental}.

Having described our adaptive SR/LR approach, we now compare it with the fixed-cutoff network. Both models follow the same two-stage training protocol. We first generate 2400 perturbed He--benzene configurations and label them with the \texttt{Orb} teacher, selected from the teacher comparison in Table~\ref{table1}, to distill the student model.
A separate set of 2400 CCSD(T)/CBS reference configurations from Ref.~\cite{akram2026} are then used for fine-tuning, split evenly into training and validation subsets.  
The fixed-cutoff and SAPT-adaptive SR/LR models are fine-tuned on the same 50\% training set and evaluated on the same held-out 50\% validation set, so the reported MAEs reflect validation errors. Detailed cutoff choices, loss terms, and training hyperparameters are given in the Supplemental Material~\cite{supplemental}.

Figure~\ref{models} compares the fixed-cutoff and SAPT-adaptive SR/LR models against CCSD(T)/CBS reference energies. 
\begin{figure}[t]
 \centering
 \includegraphics[width=1\linewidth]{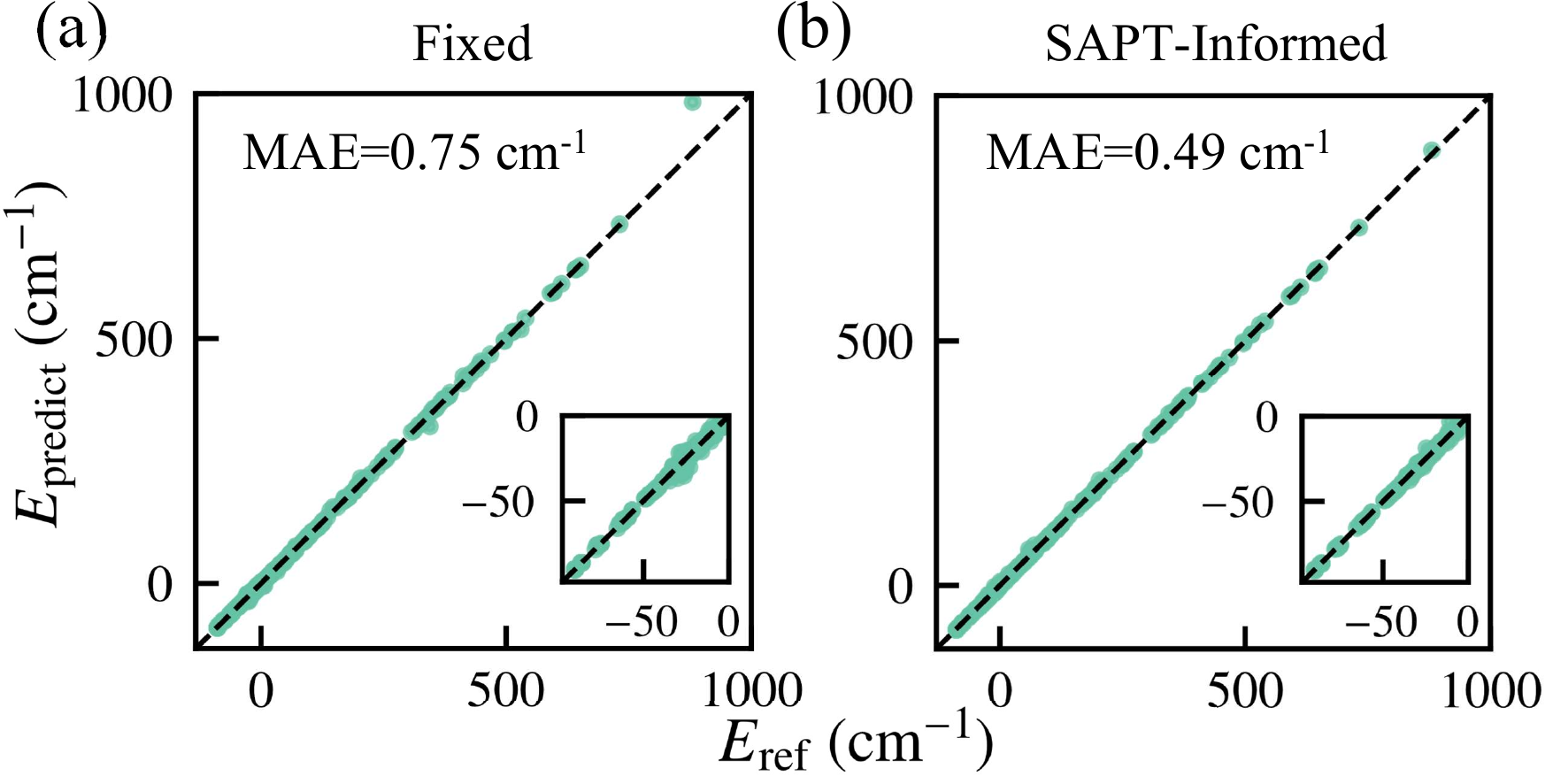}
 \caption{{Predicted versus reference binding energies for He--benzene.}
     Comparison of fixed-cutoff~\cite{ji2025machine} and SAPT-adaptive SR/LR models on the held-out validation set. The full CCSD(T)/CBS dataset contains 2400 configurations and is split evenly into training and validation subsets. 
 (a) The fixed-cutoff SR/LR model gives an MAE of \SI{0.75}{\centi\meter^{-1}}. 
 (b) The SAPT-adaptive model, which uses atom-wise SR cutoffs derived from the SAPT-informed cutoff predictor, reduces the MAE to \SI{0.49}{\centi\meter^{-1}}. 
Insets show the attractive region near the He--benzene equilibrium geometry, where the improvement is most visible.}
 \label{models}
\end{figure}
The fixed-cutoff model reaches an MAE of \SI{0.75}{\centi\meter^{-1}} and \SI{0.058}{\centi\meter^{-1}}/atom, while the SAPT-adaptive model reduces the error to \SI{0.49}{\centi\meter^{-1}} and \SI{0.038}{\centi\meter^{-1}}/atom. Rather than being uniform over the full energy range, this improvement is concentrated in the attractive region near the binding well---where accurate relative energies are particularly important for adsorption behavior---as highlighted by the insets.

Next, we investigated the effect of teacher selection on the quality of the distilled student models. To this end, we considered 18 general-purpose MLIPs as candidate teachers. For each teacher, labels were generated for the He--benzene dataset, and an identical student architecture was subsequently trained using the corresponding teacher-labeled data. The resulting student models were then evaluated against CCSD(T) reference energies on the validation set. Consequently, the errors reported in Table~\ref{table:teacher_validation_compact} reflect the performance of the distilled student models obtained from different teachers rather than the direct prediction errors of the teacher models themselves. The results demonstrate that teacher selection has a substantial impact on the accuracy of the distilled student. Among all teachers considered, \texttt{Orb} and \texttt{MatterSim} yielded the lowest validation MAEs. Accordingly, these two models were selected as teachers for the subsequent studies.
\begin{table}[h]
\centering
\caption{Mean absolute errors (in \si{\centi\meter^{-1}}) of different distilled models before and after CCSD(T) fine-tuning. Further details are provided in the supplement \cite{supplemental}.}
\begin{tabular}{l r r}
\toprule
Model & \phantom{-------} Before & \phantom{-------} After\\
\midrule
\texttt{Orb-v3-OMol}~\cite{neumann2024orb}                       & 50.55    & 0.49 \\
\texttt{M3GNet}~\cite{chen2022universal,deng2023chgnet}          & 155.29   & 3.42 \\
\texttt{Allegro-OAM-L}~\cite{Musaelian2023}                      & 221.01   & 4.04 \\
\texttt{MatterSim-v1.0.0-1M}~\cite{yang2024mattersim}            & 316.61   & 0.88 \\
\texttt{MACE-MPA-0}~\cite{batatia2022mace,batatia2025foundation} & 352.76   & 1.28 \\
\texttt{DPA-3.1-3M-FT}~\cite{zhang2026dpa3}                      & 408.03   & 1.17 \\
\texttt{Nequip-OAM-XL}~\cite{batzner2022e3nn}                    & 747.36   & 1.08 \\
\texttt{CHGNet}~\cite{Deng2023}                                  & 781.99   & 1.33 \\
\texttt{eSEN-30M-OAM}~\cite{pmlr-v267-fu25h}                     & 797.08   & 4.57 \\
\texttt{UMA-S-1P1}~\cite{wood2026umafamilyuniversalmodels}       & 1330.75  & 13.52 \\
\texttt{MatRIS\_v0.5.0\_MPtrj}~\cite{zhou2026}                   & 1489.24  & 2.81 \\
\texttt{PET-OAM-XL}~\cite{bigi2026}                              & 1635.27  & 2.39 \\
\texttt{AlphaNet-v1-OAM}~\cite{Yin2025}                          & 2028.76  & 6.10 \\
\texttt{SevenNet-MF-ompa}~\cite{kim_sevennet_mf_2024}            & 2715.94  & 2.20 \\
\texttt{DPA-4.0-Pro-MPtrj}~\cite{li2026dpa4}                     & 5171.56  & 3.88 \\
\texttt{HIENet}~\cite{yan2025}                                   & 12403.69 & 2.89 \\
\texttt{GRACE-2L-OAM-L}~\cite{Lysogorskiy2026}                   & 16604.86 & 27.29 \\
\texttt{eqV2\_M}~\cite{equiformer_v2}                            & 18834.62 & 4.22 \\
\bottomrule
\end{tabular}
\label{table:teacher_validation_compact}
\end{table}

\section{Data Efficiency and Transferability}

Having established the teacher-guided distillation workflow and the SAPT-informed SR/LR student architecture, we now test how the resulting models perform in two settings that probe different aspects of generalization. First, we quantify data efficiency on the He--benzene benchmark by comparing direct CCSD(T) training, DFT pre-training followed by CCSD(T) fine-tuning, and MLIP-guided distillation followed by CCSD(T) fine-tuning using the SAPT-informed SR/LR student model. This test asks how much high-fidelity CCSD(T) data are needed to reach a given validation error. Second, we examine transfer beyond He--benzene using a circumarene series of larger PAHs. In these cases, CCSD(T) and SAPT data are not available, and we use a fixed-cutoff SR/LR model, comparing two MLIP teachers to isolate how the teacher used during distillation affects the DFT-fine-tuned model across related molecular sizes.

\subsection{Efficiency of the Training Data Set}

\begin{figure}[t]
\centering
\includegraphics[width=1\linewidth]{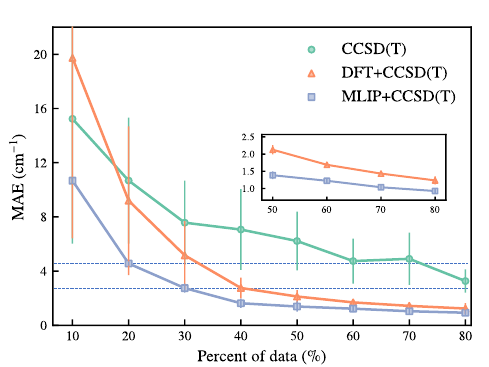}
\caption{{Data efficiency of three training routes for He--benzene.} Mean absolute error (MAE) as a function of the percentage of CCSD(T) data used in the final supervised stage. The percentages refer to fractions of the original $80\%$ training pool; all models are evaluated on the same held-out $20\%$ validation set. Points show the mean over 12 independent training runs, and error bars show the standard deviation. The direct CCSD(T) route is shown in green, the DFT+CCSD(T) route in orange, and the MLIP+CCSD(T) route in blue. The inset magnifies the high-data regime from $60$ to $80\%$; dashed horizontal lines provide reference MLIP+CCSD(T) errors for the comparison discussed in the main text.}
\label{efficiency}
\end{figure}

We evaluate the SAPT-informed SR/LR student model in the low-data regime, where the computational cost of CCSD(T) reference calculations is the practical bottleneck and the relative value of different pretrained starting points is most visible. Figure~\ref{efficiency} shows the MAE as a function of the percentage of CCSD(T) data used in the final supervised stage for three training routes: direct CCSD(T) training, DFT pre-training followed by CCSD(T) fine-tuning, and MLIP-guided distillation followed by CCSD(T) fine-tuning.
For each training fraction (\% of data), we repeat the experiment 12 times using different random subsets of the original CCSD(T) training pool, while evaluating all models on the same held-out validation set. The plotted points show the mean validation MAE across these repeats, with error bars showing one standard deviation.

Across the data range considered, pre-training plus fine-tuning outperform direct CCSD(T) training. The improvement is largest in the low-data regime. With only $20\%$ of the CCSD(T) data, the MLIP+CCSD(T) methodology already reaches a mean MAE of \SI{4.56}{\centi\meter^{-1}}, comparable to direct CCSD(T) training with $60\%$ of the data (\SI{4.74}{\centi\meter^{-1}}). At $30\%$, the MLIP+CCSD(T) mean MAE of \SI{2.74}{\centi\meter^{-1}} falls below direct CCSD(T) training even at the full $80\%$ fraction (\SI{3.26}{\centi\meter^{-1}}). Thus, the teacher-guided route substantially reduces the amount of high-fidelity data required to reach a given error level, although the reduction should be read from the full trend rather than from a single crossing point. On a standard high-performance computing cluster, this corresponds to a reduction of approximately $5\times10^5$ CPU hours, or about $63\%$, in the compute budget needed to generate the CCSD(T) training data. 

The direct CCSD(T) curve is not strictly monotonic; for example, the mean error at $70\%$ is slightly higher than at $60\%$. We do not interpret this small non-monotonicity as a physical effect, but rather as a consequence of stochastic variation across finite training splits and validation statistics. Despite this variation, the pretrained routes consistently reach lower MAEs with smaller CCSD(T) fractions, supporting the conclusion that they reduce the amount of high-fidelity data required to reach the same accuracy.

The two pre-training routes are closest at large CCSD(T) fractions, as shown in the inset of Fig.~\ref{efficiency}. In the $60$--$80\%$ range, both DFT+CCSD(T) and MLIP+CCSD(T) achieve sub-\SI{2}{\centi\meter^{-1}} errors, but the MLIP+CCSD(T) route remains consistently lower. At the smallest data fractions, the MLIP+CCSD(T) route also shows smaller uncertainty than DFT+CCSD(T), suggesting a more stable starting point when high-fidelity supervision is most limited. The MLIP route has an additional practical advantage: unlike DFT pre-training, which requires explicit electronic-structure calculations for each configuration~\cite{burke2012perspective}, MLIP labels are generated by evaluating an already-trained force field, making large-scale teacher labeling substantially less costly. After fine-tuning, the specialized student model is also much cheaper to evaluate than the teacher itself. For the same set of 1000 He--benzene configurations on the same CPU setup, the \texttt{Orb} teacher requires \SI{65.39}{\second}, whereas the SAPT-informed student model requires \SI{2.34}{\second}, giving a speedup of about $28\times$. The reduction is consistent with the much smaller model size: the student contains only $4.25\times10^{5}$ trainable parameters, compared with approximately $2.55\times10^{7}$ for \texttt{Orb}. While the inference times above do not account for the cost of generating CCSD(T) or SAPT data, treating these as sunk high-performance computing costs yields substantial deployment advantages for the more compact fine-tuned model on resource-constrained hardware.

A small CCSD(T) dataset is more effective when it is used to correct a reasonable interaction surface produced from a general MLIP than when it must determine the surface without a pretrained starting point. This point is especially important for He--benzene, where the interaction surface is shallow, anisotropic, and shaped by a geometry-dependent SR/LR crossover. In the MLIP+CCSD(T) route, the student is first trained on a large set of MLIP-labeled, target-relevant configurations; CCSD(T) fine-tuning then corrects this initial model with high-fidelity data. The comparison with DFT+CCSD(T) suggests that the pretrained teacher matters: even when the same student architecture and CCSD(T) fine-tuning protocol are used, different label sources lead to different final errors and uncertainties. 

\subsection{Transferability to Larger Arenes}

\begin{figure*}[t]
    \centering
    \includegraphics[width=1\linewidth]{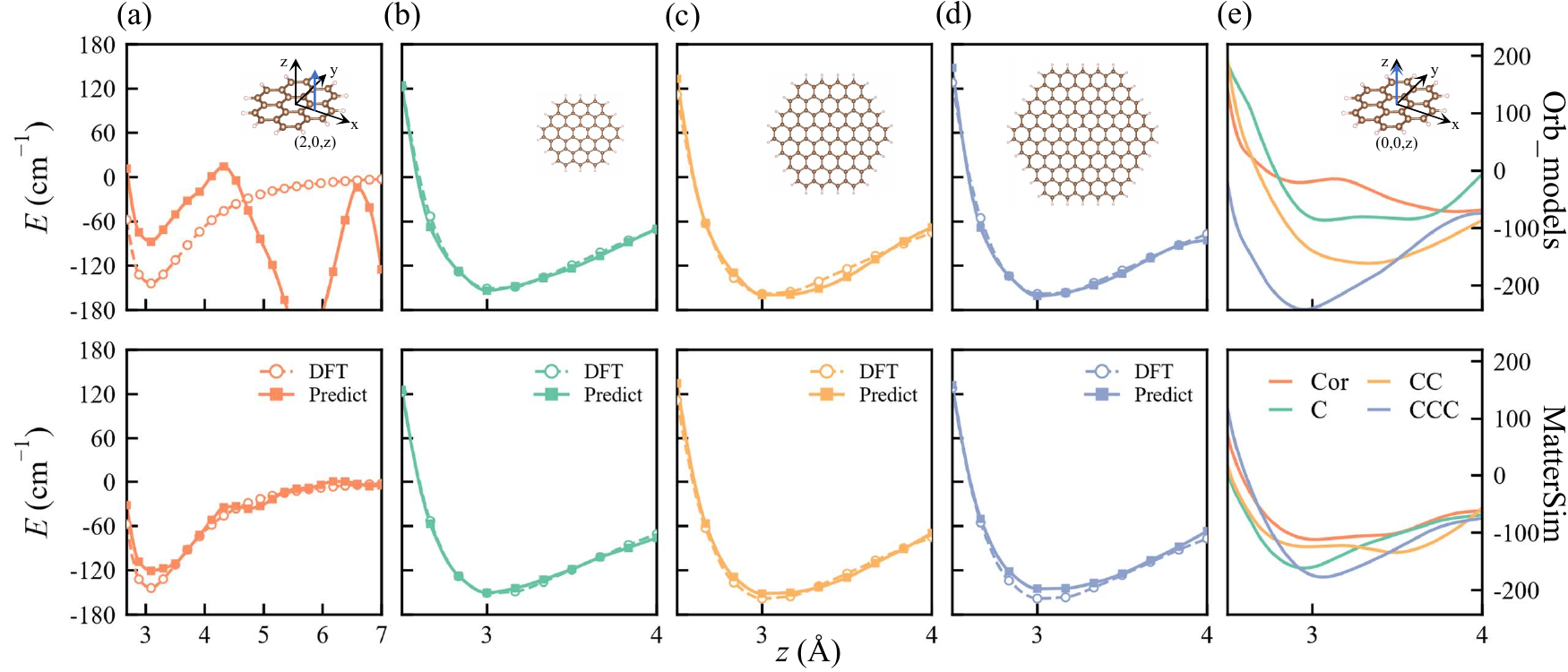}

    \caption{{Teacher-conditioned transferability across a graphene-like structural series.}
    Predicted He--PAH interaction profiles for coronene (a), C-coronene (b), CC-coronene (c), and CCC-coronene (d) using a single fixed-cutoff SR/LR model trained on the combined four-molecule dataset. 
    The model is first distilled from either an \texttt{Orb} teacher (top row) or a \texttt{MatterSim} teacher (bottom row), and is then fine-tuned to DFT reference energies. 
    In panels (a)--(d), which all share a common $y$-axis, open symbols with dashed lines denote DFT reference data and closed symbols with solid lines show the corresponding interpolated model predictions along the off-center out-of-plane trajectory $(2,0,z)$ (blue arrow in (a) inset top panel). 
    Panel (e) compares the same trained models evaluated along the high-symmetry central-axis trajectory $(0,0,z)$ for the four PAHs (blue arrow in (e) inset top panel) where there is limited training data. Changing the teacher mainly affects the coronene profile, while the larger PAHs remain comparatively stable.
}
    \label{transfer}
\end{figure*}

To test how teacher choice affects specialization beyond the He--benzene benchmark, we consider a circumarene series of PAHs of increasing size: coronene, circumcoronene (C-coronene), circumcircumcoronene (CC-coronene), and circumcircumcircumcoronene (CCC-coronene)~\cite{lazar2013adsorption}. For this test, we train one fixed-cutoff SR/LR model on the combined data from all four PAH systems, rather than training separate models for each molecule. The model follows the same two-stage procedure used above: it is first distilled from teacher-labeled configurations and then fine-tuned using DFT interaction energies as the high-fidelity training data for the combined PAH dataset. Details of the DFT calculations and data generation are provided in the Supplemental Material~\cite{supplemental}. Because no CCSD(T) reference data are available for these larger PAHs, the DFT data are used as the high-fidelity reference for this transferability test.

We use the fixed-cutoff SR/LR model because generating SAPT decompositions for coronene and larger PAHs would be substantially more expensive, so the comparison focuses on the effect of the teacher used during distillation while keeping the student architecture fixed. The MAEs reported below are evaluated only along the one-dimensional off-center out-of-plane cuts shown in Figs.~\ref{transfer}(a)--(d), not over the full configuration space. These cuts correspond to representative He--PAH approach geometries above the molecular plane. For the PAH series, the evaluation trajectory is taken at a fixed lateral position $(x,y)=(2,0)~\mathrm{\AA}$, with the closest available lateral position used for coronene to match the same trajectory as closely as possible. The reference curves in Figs.~\ref{transfer}(a)--(d) are DFT reference energies along these $(2,0,z)$ cuts, and the model predictions are evaluated on the same trajectories. Figure~\ref{transfer}(e) provides an additional comparison of model predictions along the high-symmetry central-axis trajectory, $(0,0,z)$, illustrating the evolution of the interaction profile with molecular size.

For the off-center $(2,0,z)$ cuts shown in Figs.~\ref{transfer}(a)--(d), the \texttt{Orb}-based teacher gives low errors for the larger PAHs, with MAEs of \SIlist{0.046;0.057;0.030}{\centi\meter^{-1}}/atom for C-coronene, CC-coronene, and CCC-coronene, respectively. Its performance is substantially worse for bare coronene, where the MAE increases to \SI{2.26}{\centi\meter^{-1}}/atom and the largest deviations occur near the attractive well and the repulsive-to-attractive crossover region. Replacing the \texttt{Orb}-based teacher with \texttt{MatterSim} reduces the coronene MAE from \SI{2.26}{\centi\meter^{-1}}/atom to \SI{0.200}{\centi\meter^{-1}}/atom, a factor of about $11$. For the larger PAHs, the two teachers give comparable errors, with the \texttt{MatterSim}-based model giving MAEs of \SIlist{0.035;0.054;0.040}{\centi\meter^{-1}}/atom for C-coronene, CC-coronene, and CCC-coronene, respectively. The central-axis comparison in Fig.~\ref{transfer}(e) shows the same qualitative trend: the teacher choice mainly affects the smaller coronene system, while the profiles for the larger PAHs remain comparatively stable. Since the student architecture, DFT fine-tuning data, and evaluation trajectories are otherwise fixed, the difference between the two rows reflects the effect of the teacher used during distillation.

The two teacher models differ in more than their average accuracy on the PAH cuts. They also lead to different error patterns across the structural series. With the \texttt{Orb}-based teacher, the combined PAH model is highly accurate for C-coronene, CC-coronene, and CCC-coronene, but shows a much larger error for bare coronene, especially near the binding well. With the \texttt{MatterSim}-based teacher, the coronene error is strongly reduced while the errors for the larger PAHs remain at a similar level. This suggests that the teacher dependence is not a uniform shift in accuracy across the series, but is concentrated in how the model describes the smallest PAH environment. One possible explanation is that the distilled representation encodes different balances among short-range repulsion, weak attractive binding, and geometric anisotropy as the adsorption environment changes with molecular size. Figure~\ref{transfer}(e) provides a complementary view by evaluating the same trained model along the central $z$ axis for all four structures, showing how the predicted interaction profile evolves toward the larger graphitic limit.

\section{Discussion}

The results presented here support hybrid MLIP--CCSD(T) distillation as a practical route to quantum-chemical accuracy with minimal high-fidelity supervision. In this picture, a pretrained machine-learning interatomic potential supplies the student with a useful starting point, and a small high-fidelity CCSD(T) dataset refines that starting point at the target level of theory. For the He--benzene benchmark, this data-efficient adaptation is further strengthened by the SAPT-informed adaptive SR/LR architecture, which uses the geometry-dependent short-range/long-range crossover to improve the description of the attractive binding region and reduce the validation MAE from \SI{0.75}{\centi\meter^{-1}} to \SI{0.49}{\centi\meter^{-1}}. Beyond reducing the amount of coupled-cluster data required for training, the specialized student also provides a substantially more compact and faster surrogate at inference time: on the same CPU setup, it evaluates 1000 He--benzene configurations about $28\times$ faster than the \texttt{Orb} teacher. We find that MLIP-guided distillation substantially reduces the number of coupled-cluster labels needed to reach a given CCSD(T)-level error: in the He--benzene benchmark, $30\%$ of the CCSD(T) data after MLIP distillation outperforms direct CCSD(T) training using the full $80\%$ training fraction. The transferability tests further show that the choice of MLIP teacher can influence how the specialized model behaves across a structural family. The broader teacher sweep in Table~S2 in the Supplemental Material~\cite{supplemental} further supports the generality of the strategy by showing that multiple pretrained teachers can be adapted through the same CCSD(T) fine-tuning protocol, while also demonstrating that teacher compatibility cannot be inferred from the distilled-student error alone.

The results also identify the current scope of the introduced framework. First, it would benefit from a more systematic procedure for specialization. At present, identifying an effective student architecture, training protocol, and hyperparameter set can require substantial tuning, which limits immediate high-throughput deployment. The SAPT-informed model also relies on SAPT data, which can be costly to generate (21.08 CPU hours for a single point calculation for He--benzene system), especially for larger systems. This affects the scalability of the present implementation, although the same strategy can be adapted to whatever level of theory is available. Second, our results show that teacher identity plays an important role, but a predictive theory of teacher compatibility remains an open opportunity. In the present study, choosing an appropriate MLIP still relies on physical intuition and prior experience with the strengths and biases of different teachers. The first challenge is largely practical and can be addressed through automation. The second is more fundamental, since it requires a deeper understanding of how pretrained representations align with different target interaction classes. These considerations define the present utility of the framework while leaving the principal conclusion unchanged: specialization can substantially improve the performance of MLIPs for targeted non-covalent interaction problems.

These observations suggest several specific directions for future work. One is to make specialization more efficient and systematic through improved hyperparameter selection, more robust student-design strategies, and less dependence on expensive auxiliary data such as SAPT for defining adaptive partitions. Another is to develop quantitative measures of teacher compatibility, for example from chemical-space overlap, interaction decomposition analysis, or representation-space similarity. Such tools could help evaluate whether a teacher is suitable for a target problem before extensive downstream training. It would also be useful to extend the framework to richer multi-fidelity schemes, where corrections are introduced gradually across levels of theory~\cite{zaspel2018boosting, ramakrishnan2015big}, and to combine it with active learning so that expensive reference data are added only where the transferred prior remains inadequate~\cite{smith2020ani, vandermause2020fly,zhang2019active,uteva2018active}. 

In conclusion, we have developed a hybrid distillation and fine-tuning framework for learning weak non-covalent intermolecular interactions with reduced high-fidelity data requirements. For the He--benzene system considered here, MLIP-guided distillation followed by CCSD(T) fine-tuning reaches sub-\SI{4}{\centi\meter^{-1}} accuracy with substantially fewer coupled-cluster labels than direct CCSD(T) training: using $20\%$ of the CCSD(T) data gives an error comparable to direct training with $60\%$, while $30\%$ outperforms direct training with the full $80\%$ training fraction. A SAPT-informed adaptive short-range/long-range architecture, in which the short-range cutoff is determined dynamically from the interaction geometry, further lowers the validation MAE from \SI{0.75}{\centi\meter^{-1}} to \SI{0.49}{\centi\meter^{-1}} by capturing the anisotropic crossover between short-range repulsion and long-range dispersion. Across a graphene-like series of polycyclic adsorbates, changing the MLIP teacher alters the coronene error by an order of magnitude while leaving the larger PAHs comparatively stable, showing that teacher choice can affect transfer behavior even when the architecture and fine-tuning protocol are fixed.
These results point to hybrid distillation and fine-tuning as a practical strategy for building accurate potentials for weak intermolecular interactions from broad pretrained models and a small number of targeted high-fidelity reference calculations.

\section*{Data and Code Availability}
The source code and full dataset associated with this work are publicly available \cite{paperrepo}.

\begin{acknowledgments}
This work was primarily supported by the National Science Foundation Materials Research Science and Engineering Center program through the UT Knoxville Center for Advanced Materials and Manufacturing~(DMR-2309083).
\end{acknowledgments}

\bibliography{ref}

\ifarXiv
\clearpage
\onecolumngrid


\section*{Supplementary material}

# Supplemental Material for "Non-covalent Interactions at $cm^{-1}$ Accuracy: Data Efficient Physics-Informed Distillation for Machine Learning Interatomic Potentials"

Yulin Shen,[1] Shahzad Akram,[2] Louis Primeau,[1] Gen Zu,[3] Konstantinos D. Vogiatzis,[2] Yang Zhang,[1,4] and Adrian Del Maestro[1,4]

[1]*Department of Physics and Astronomy, University of Tennessee, Knoxville, Tennessee 37996, USA*
[2]*Department of Chemistry, University of Tennessee, Knoxville, TN, 37996, USA*
[3]*Max Planck Institute for Chemical Physics of Solids, 01187, Dresden, Germany*
[4]*Min H. Kao Department of Electrical Engineering and Computer Science, University of Tennessee, Knoxville, Tennessee 37996, USA*

## I. REFERENCE ELECTRONIC-STRUCTURE DATA

The high-fidelity reference data for the He–benzene system were taken from our previous study, in which a benchmark potential-energy surface was constructed for the helium–benzene interaction [1]. That work provides both coupled-cluster reference energies and symmetry-adapted perturbation theory (SAPT) energy decompositions. The coupled-cluster data were computed at the CCSD(T)/CBS level, where CCSD(T) denotes coupled-cluster theory with single and double excitations plus a perturbative treatment of triple excitations, and CBS denotes extrapolation to the complete-basis-set limit. This level of theory is commonly used as a high-accuracy reference for weak non-covalent interactions, but its computational cost limits the number and size of systems that can be treated directly.

The SAPT calculations from the same study were used to analyze the physical origin of the He–benzene interaction. SAPT decomposes the interaction energy into electrostatic, exchange-repulsion, induction, and dispersion contributions. In the present work, this decomposition is used to construct the direction-dependent short-range/long-range crossover radius for He–benzene. The detailed electronic-structure settings used for the CCSD(T)/CBS and SAPT calculations are given in Ref. [1].

For the larger polycyclic aromatic hydrocarbons (PAHs), direct CCSD(T)/CBS calculations are not practical because of the much larger system size. We therefore used density functional theory (DFT) interaction energies as the reference data for He–coronene, He–circumcoronene (C-coronene), He–circumcircumcoronene (CC-coronene), and He–circumcircumcircumcoronene (CCC-coronene). Specifically, we used the PBE0-D4/def2-SVP level of theory, where PBE0 is a hybrid exchange–correlation functional, D4 is an empirical dispersion correction, and def2-SVP is the atomic-orbital basis set. This choice follows the benchmarking in Ref. [1], where PBE0-D4/def2-SVP was shown to reproduce the coupled-cluster He–benzene reference data with an MAE of 3.29 $cm^{-1}$ along representative interaction cuts. Because the larger PAHs preserve the same local helium–$\pi$ interaction motif as benzene, we use this validated DFT level to generate the interaction-energy data for the PAH transferability tests.

## II. NEURAL NETWORK ARCHITECTURE

### II.1. Neural Network from DeepMD

Before using the SR/LR architecture, we first tested a standard multilayer-perceptron neural network implemented in the DeepMD framework [2–4]. This model was used as a local-descriptor baseline, in which the binding energy is fitted directly from the atomic species and Cartesian coordinates of each He–benzene configuration, without an explicit short-range/long-range decomposition.

The model uses the DeepMD `se_e2_a` descriptor to construct local atomic environments. Neighbor atoms are included within a cutoff radius of 8.0 Å, with a smoothing distance of 0.5 Å. The neighbor-selection parameter was set to 32 for each atomic species in the type map. The descriptor embedding network has hidden-layer dimensions `[32, 64, 128]`, and the number of axis neurons is 24.

The descriptor output is passed to a fitting network that predicts atomic energy contributions. The fitting network is a fully connected MLP with hidden-layer dimensions `[240, 240, 240]` and `tanh` activation. The total binding energy is obtained by summing the atomic energy contributions over all atoms.

The model was trained using an energy-only loss, with the force and virial loss prefactors set to zero. The energy prefactor was increased from 1 to 2 during training. Training was performed for 1.6 million optimization steps using an exponential learning-rate schedule, starting from $1.0 \times 10^{-4}$, with a decay step of 80,000 and a final learning rate of $5.0 \times 10^{-5}$. The result for this model is plotted in Fig. S1.

### II.2. Short-Range/Long-Range Neural Network

We implemented two SR/LR neural-network models: a fixed-cutoff SR/LR model and a SAPT-informed adaptive-cutoff SR/LR model. Both models take the atomic species, Cartesian coordinates, and simulation box as input. The He atom index is identified from the atomic type map, and all radial descriptors are constructed using He-centered distances. Both models output the total binding energy as the sum of a short-range contribution and a long-range contribution.

In the fixed-cutoff SR/LR model, the local descriptor is constructed using two radial shells with cutoff radii `[1.7, 8.0]` Å. Each shell contains 28 Gaussian radial basis functions, so the descriptor dimension for each atom is $2 \times 28 = 56$. The radial basis centers are uniformly distributed from zero to the corresponding shell cutoff. A cosine cutoff function is applied within each shell.

The SR branch maps the 56-dimensional descriptor to atomic short-range energy contributions. The SR fitting network is a fully connected MLP with hidden-layer dimensions `[576, 576]` and `SiLU` activation. The scalar output of this network gives the atomic SR energy, and the total SR energy is obtained by summing over all atoms.

The LR branch in the fixed-cutoff model uses the same 56-dimensional radial descriptor. This descriptor is passed to a latent network with dimensions `[56, 576, 8]`, using `SiLU` activation after the first linear layer. The resulting 8-dimensional latent vector for each atom is used as the input to a reciprocal-space SOG block. The SOG block contains 8 layers and 12 Gaussian functions per layer. Reciprocal vectors are generated from integer triplets $(n_1, n_2, n_3)$ with each component in the range $[-2, 2]$, excluding the zero vector. This gives 124 reciprocal vectors. The Gaussian centers are initialized on a logarithmic grid between $10^{-3}$ and 5.0 in reciprocal-space magnitude, while the Gaussian widths and weights are trainable. The LR energy is computed from the squared magnitude of the latent reciprocal density weighted by the learned SOG kernel.

The SAPT-informed SR/LR model uses the same overall SR/LR structure, but replaces the fixed SR cutoff by an atom-wise adaptive cutoff. For each configuration, a SAPT-derived center cutoff is first obtained from a spherical-harmonic fit stored in `rc_fit.npz`. The direction from the benzene center to the He atom is converted to spherical angles, and the center cutoff is evaluated from the real spherical-harmonic expansion. This value is clipped to the range 0.2–20.0 Å before being used in the training loss.

The adaptive SR cutoff is predicted by a pair cutoff network. Each atomic species is embedded into a 32-dimensional trainable vector. For each atom, this embedding is concatenated with the He–atom unit vector and the He–atom distance, giving an input dimension of $32 + 3 + 1 = 36$. The cutoff predictor is an MLP with hidden-layer dimensions `[128, 128]` and `SiLU` activation. The scalar output is clipped to the range 0.8–12.0 Å. The predicted cutoff for the He atom itself is fixed to 1.0.

TABLE S1: Mean absolute error (MAE, in $\mathrm{cm}^{-1}$) of He–benzene binding energies from three NNs with different architectures.

| Model | MAE ($\mathrm{cm}^{-1}$) |
|---|---|
| DeepMD | 6.44 |
| Fixed SR/LR | 0.75 |
| SAPT-informed SR/LR | 0.49 |

Using the predicted atom-wise SR cutoffs, the adaptive SR descriptor is constructed with 28 Gaussian radial basis functions and two scaled radial shells `[0.3, 1.0]`. The radial coordinate is normalized by the product of the shell scale and the predicted atom-wise cutoff. The descriptor dimension is again $2 \times 28 = 56$. Values outside the corresponding scaled cutoff are masked by a compact-support cutoff. The adaptive SR descriptor is then passed to the same SR fitting network structure as above, with hidden-layer dimensions `[576, 576]` and `SiLU` activation.

For the LR branch of the SAPT-informed model, the descriptor uses a fixed LR cutoff. In the final training script, this cutoff was set to 12.0 Å. The LR descriptor again has dimension 56 and is passed through a latent network with dimensions `[56, 576, 8]`. The resulting latent coefficients are evaluated by the same reciprocal-space SOG block, with 8 layers, 12 Gaussian functions per layer, and reciprocal vectors generated from integer triplets in the range $[-2, 2]$ excluding the zero vector.

Both SR/LR models were trained with energy and force losses. The energy loss is the mean-squared error between the predicted and reference binding energies. Forces are obtained by automatic differentiation of the predicted total energy with respect to the atomic coordinates, and the force loss is included when force labels are available. In the final training scripts, the force prefactor was set to 0.005. The models were optimized with AdamW using a learning rate of $1.0 \times 10^{-4}$ and weight decay $1.0 \times 10^{-4}$. A step learning-rate scheduler with step size 8000 and decay factor 0.5 was used. Training was performed for 10,000 epochs with batch size 64.

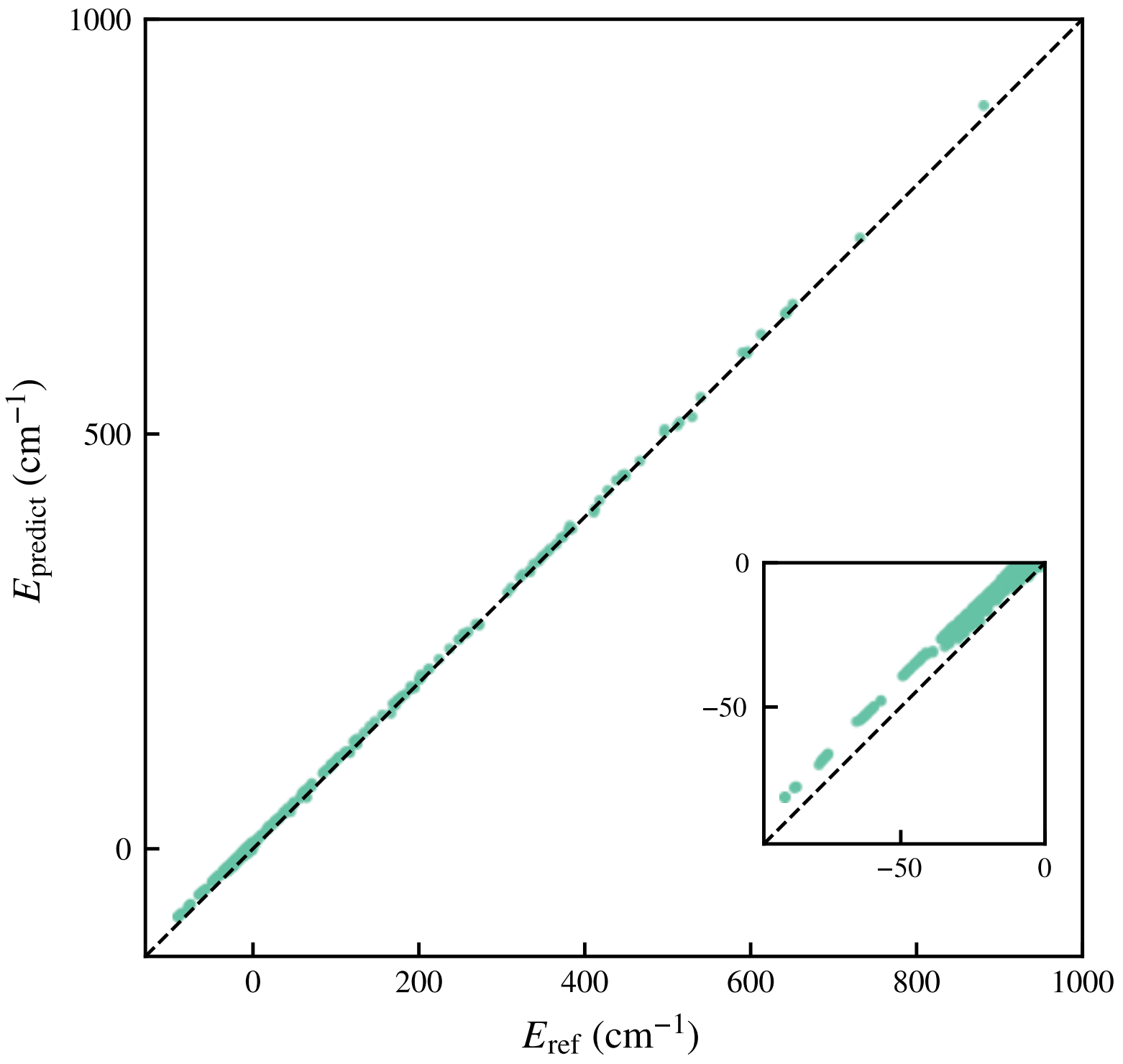


FIG. S1: Predicted versus reference binding energies for He–benzene interactions using NN from DeepMD.

For the SAPT-informed model, two additional cutoff regularization terms were included. The first penalizes the variance of the predicted cutoffs among equivalent C and H atoms. The second aligns the mean predicted cutoff for C and H atoms with the SAPT-derived center cutoff. The variance-penalty weight was annealed from 1.0 to 0.5, and the center-alignment weight was annealed from 0.2 to 0.05. Both annealing schedules start at 30% of the total training epochs and end at 80%. The energy/force loss, cutoff-variance loss, and center-alignment loss were combined using trainable uncertainty weights. The comparison for these three models is listed in Table. S1.

## III. SYMMETRY-ADAPTED PERTURBATION THEORY-INFORMED SHORT-RANGE CUTOFF

The SAPT-informed short-range cutoff was constructed in two steps. First, we extracted a direction-dependent crossover radius from the SAPT energy decomposition. Second, the resulting cutoff map was fitted by a smooth spherical-harmonic expansion so that it could be evaluated continuously during neural-network training.

For each SAPT configuration, the He position relative to the benzene center was written in spherical coordinates $(R, \theta, \phi)$, where $\theta$ is the polar angle measured from the positive $z$ direction and $\phi$ is the azimuthal angle in the molecular plane. The SAPT energy components were decomposed into electrostatics, exchange, induction, and dispersion contributions (Fig. S2(a)). For each direction $(\theta, \phi)$, we defined the short-range/long-range crossover radius by the condition

$$|E_{\mathrm{elst}} + E_{\mathrm{exch}} + E_{\mathrm{ind}}| = |E_{\mathrm{disp}}|.$$

Here the left-hand side contains the non-dispersion SAPT contributions, including the short-range exchange-dominated part, while the right-hand side represents the dispersion contribution.

Because the SAPT data were available only on a discrete set of Cartesian He positions, the four SAPT components were first interpolated using a radial-basis-function interpolator. We used a thin-plate-spline RBF interpolator with smoothing parameter $10^{-6}$ and up to 200 neighboring points. For a given direction $(\theta, \phi)$, the function

$$g(R; \theta, \phi) = |E_{\mathrm{elst}}(R, \theta, \phi) + E_{\mathrm{exch}}(R, \theta, \phi) + E_{\mathrm{ind}}(R, \theta, \phi)| - |E_{\mathrm{disp}}(R, \theta, \phi)|$$

was evaluated on a radial scan. The root of this function was then found using a Brent root solver, giving the crossover radius $R_{\mathrm{cut}}(\theta, \phi)$.

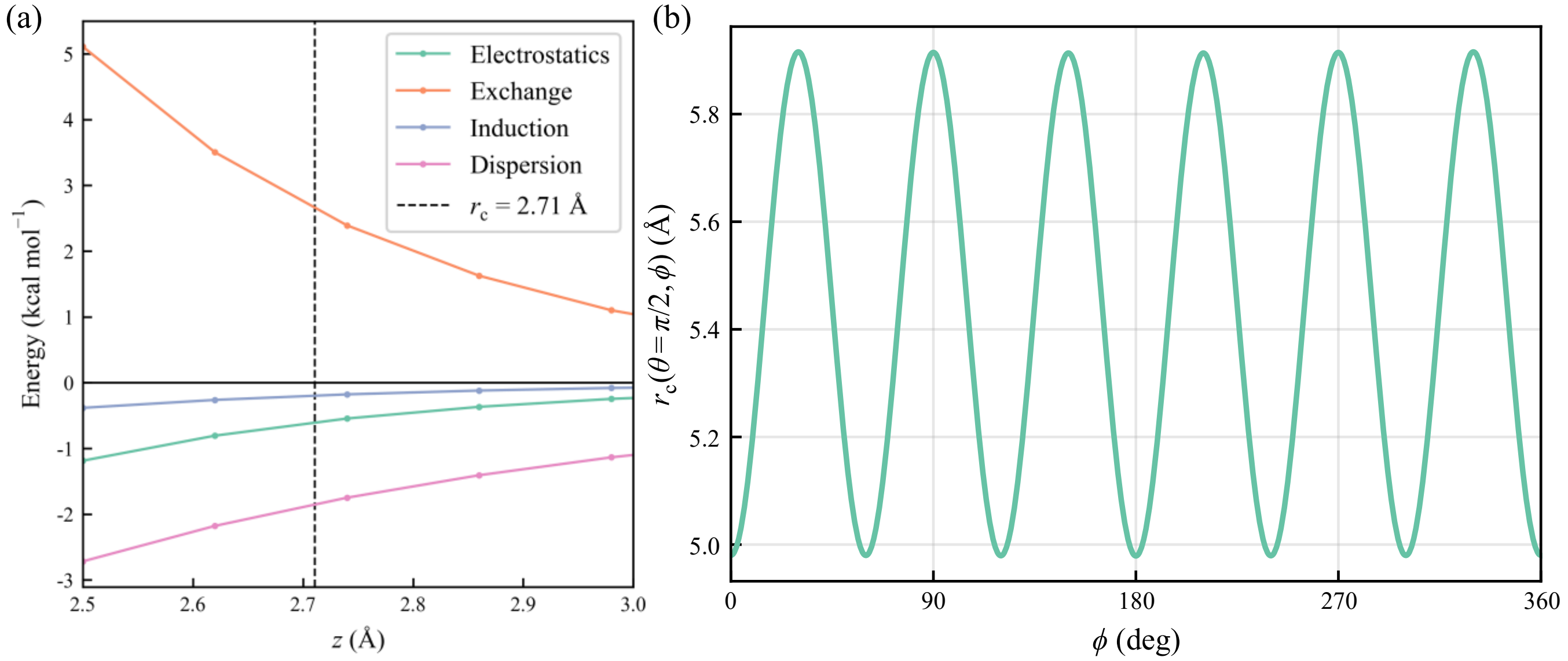


FIG. S2: **Construction of the SAPT-informed short-range cutoff.** (a) Representative SAPT energy decomposition along a selected He approach direction. The crossover radius $r_c$ is defined by the condition $|E_{\mathrm{elst}} + E_{\mathrm{exch}} + E_{\mathrm{ind}}| = |E_{\mathrm{disp}}|$, which separates the short-range-dominated and dispersion-dominated regimes. (b) The crossover radius $r_c$ on the $z = 0$ cut, showing a sixfold angular modulation consistent with $C_6$ symmetry.

The root search was performed on a base angular sector and then extended to the full azimuthal range using the molecular symmetry. The base sector was chosen inside the fundamental $C_6$ wedge of benzene, with $\phi \in [0, \pi/6]$. The full $0 \le \phi < 2\pi$ range was generated by mirror symmetry within the wedge and by sixfold rotational symmetry. Mirror symmetry with respect to the molecular plane was also enforced by averaging paired values at $\theta$ and $\pi - \theta$ when both were present on the angular grid. This procedure produced a full cutoff map $R_{\mathrm{cut}}(\theta, \phi)$ on an angular grid with 80 polar-angle points and 721 azimuthal-angle points.

To obtain a smooth function for use in the neural network, the finite values of the cutoff map were fitted using real spherical harmonics. The real spherical-harmonic basis was constructed up to $l_{\max} = 10$. For each angular point, the basis contains the real $m = 0$ harmonic and the real and imaginary combinations of the complex harmonics for $m > 0$. Before fitting, mirror symmetry about the molecular plane was explicitly enforced by augmenting the data with paired points $(\pi - \theta, \phi)$ carrying the same cutoff value. The coefficients were then obtained from a linear least-squares fit to the finite cutoff values.

The fitted cutoff function can be written as

$$R_{\mathrm{cut}}^{\mathrm{fit}}(\theta, \phi) = \sum_{l=0}^{l_{\max}} \sum_{m} c_{lm} Y_{lm}^{\mathrm{real}}(\theta, \phi),$$

with $l_{\max} = 10$. This expansion contains 121 real spherical-harmonic basis functions. The fitted parameters were saved and used during training to evaluate the SAPT-derived center cutoff for each He–benzene configuration. The mean absolute error between the fitted cutoff and the original cutoff map is

$$\mathrm{MAE} = 0.001038 \text{ Å}.$$

This small fitting error indicates that the spherical-harmonic representation provides a smooth and essentially lossless approximation to the SAPT-derived cutoff surface. The resulting angular dependence is shown in Fig. S2(b), where the $z = 0$ cut, $r_c(\theta = \pi/2, \phi)$, displays a sixfold modulation consistent with the $C_6$ symmetry of benzene.

## IV. TEST ON MORE MODELS

Before choosing the teacher model used in the distillation stage, we tested several general-purpose machine-learning interatomic potentials as possible teacher-labeling routes. For each candidate teacher, we generated labels for the He–benzene dataset and trained the same student architecture using the corresponding teacher-labeled data. The


\fi

\end{document}